\documentclass[11pt,thmsa]{article}
\usepackage{amssymb}

\usepackage{graphicx}
\usepackage[reqno,intlimits]{amsmath}

\newtheorem{theorem}{Theorem}

\begin{document}

\begin{center}
\bigskip \bigskip

\bigskip

{\LARGE Matrix representation of the}

{\LARGE generalized Moyal algebra}
\end{center}

\medskip

\bigskip

\begin{center}
Jerzy F. Pleba\'{n}ski$^{\ast }\footnote{%
E-mail: pleban@fis.cinvestav.mx}$ Maciej Przanowski$^{\ast ,\ast \ast }%
\footnote{%
E-mail: przan@fis.cinvestav.mx}$

and Francisco J. Turrubiates$^{\ast }\footnote{%
E-mail: fturrub@fis.cinvestav.mx}$

\bigskip

$^{\ast }$Department of Physics

Centro de Investigaci\'{o}n y de Estudios Avanzados del IPN

Apartado Postal 14-740, M\'{e}xico, D.F., 07000, M\'{e}xico.

\smallskip

$^{\ast \ast }$Institute of Physics

Technical University of \L \'{o}d\'{z},

W\'{o}lcza\'{n}ska 219, 93-005, \L \'{o}d\'{z}, Poland.

\bigskip

\bigskip

\textbf{Abstract}
\end{center}

\bigskip

It is shown that the isomorphism between the generalized Moyal algebra and
the matrix algebra follows in a natural manner from the generalized Weyl
quantization rule and from the well known matrix representation of the
destruction and creation operators.

\bigskip

\medskip

\noindent PACS numbers: 03.65.Ca

\noindent Keywords: Deformation quantization, Quantum mechanics.

\begin{center}
\newpage
\end{center}

This short note is motivated by Merkulov's paper ``The Moyal product is the
matrix product'' [1], where the canonical isomorphism between the Moyal
algebra and an infinite matrix algebra has been found.

Here we are going to show how the results of previous works [2,3,4] and the
well known in quantum mechanichs [5,6] representation of the position, $%
\widehat{x}$, and the momentum, $\widehat{p}$, operators lead to
isomorphisms between various $\ast $-algebras and infinite matrix
algebra.\bigskip 

First remind the basic theorems [3,4].

Let $P[[x,p,\hbar ]]$ be the $\mathbb{C}$ linear space of all formal power
series of $x,p$ and $\hbar $ where $\left( x,p\right) \in \mathbb{R\times R}$
are the coordinates of the phase space $\Gamma =\mathbb{R\times R}$ and $%
\hbar $ is a real parameter (the deformation parameter). The phase space $%
\Gamma =\mathbb{R\times R}$ is endowed with usual symplectic form

\begin{equation}
\omega =dp\wedge dq  \tag{1}
\end{equation}

Let also $\widehat{P}[[\widehat{x},\widehat{p},\hbar ]]$ be an associative
algebra over $\mathbb{C}$ of the formal power series of $\widehat{x},%
\widehat{p},\hbar \widehat{1}$. The self-adjoint operators $\widehat{x}$ and 
$\widehat{p}$ act in a Hilbert space $\mathcal{H}$ and satisfy the
commutation relation 
\begin{equation}
\lbrack \widehat{x},\widehat{p}]:=\widehat{x}\widehat{p}-\widehat{p}\widehat{%
x}=i\hbar \widehat{1}.  \tag{2}
\end{equation}
As usual, $\widehat{1}$ denotes the unity operator.

$\widehat{P}[[\widehat{x},\widehat{p},\hbar ]]$ is the eveloping algebra of
the Heisenberg-Weyl algebra generated by $\widehat{x},\widehat{p},\hbar 
\widehat{1}.$

The following theorem holds [3,4]

\begin{theorem}
There exists a vector space isomorphism 
\begin{equation*}
W_{g}:P[[x,p,\hbar ]]\longrightarrow \widehat{P}[[\widehat{x},\widehat{p}%
,\hbar ]]
\end{equation*}
such that\bigskip

(i)$\quad W_{g}(1)=\widehat{1}$\bigskip

$\quad \quad W_{g}(p^{m}x^{n})=\sum\limits_{s=0}^{\min (m,n)}g(m,n,s)\hbar
^{s}\widehat{p}^{m-s}\widehat{x}^{n-s}$\bigskip

$\quad \quad m,n\in N,\quad m+n\neq 0,\quad g(m,n,s)\in \mathbb{C},\quad
g(m,n,0)=1$

\bigskip

(ii)$\quad i\hbar W_{g}\left( \left\{ x,A\right\} _{\mathcal{P}}\right) =%
\left[ \widehat{x},W_{g}(A)\right] $\bigskip

$\quad \quad i\hbar W_{g}\left( \left\{ p,A\right\} _{\mathcal{P}}\right) =%
\left[ \widehat{p},W_{g}(A)\right] $\bigskip

for every $A\in P[[x,p,\hbar ]],$ with $\left\{ \cdot ,\cdot \right\} _{%
\mathcal{P}}$ denoting the Poisson bracket.

\bigskip

Moreover, every isomorphism $W_{g}:P[[x,p,\hbar ]]\longrightarrow \widehat{P%
}[[\widehat{x},\widehat{p},\hbar ]]$ satisfies the conditions (i) and (ii)
iff 
\begin{equation}
g(m,n,s)=\frac{\left( -1\right) ^{s}m!n!}{s!\left( m-s\right) !\left(
n-s\right) !}\frac{d^{s}f(y)}{dy^{s}}|_{y=0}  \tag{3}
\end{equation}
where $f(y)=\sum\limits_{k=0}^{\infty }f_{k}y^{k},$ $f_{0}=1,$ is a formal
series independent of $\hbar \qquad \qquad \blacksquare $
\end{theorem}

(Of course, one can easily recognize in the conditions (ii) of Theorem 1,
the modified Dirac quantization rules).

Then, the second theorem reads [4]

\begin{theorem}
Let $W_{g}:P[[x,p,\hbar ]]\longrightarrow \widehat{P}[[\widehat{x},\widehat{p%
},\hbar ]]$ be the vector space isomorphism defined in Theorem 1.

Then for any $A,B\in P[[x,p,\hbar ]]$%
\begin{equation}
W_{g}\left( A\right) W_{g}\left( B\right) =W_{g}\left( A\ast _{g}B\right) 
\tag{4}
\end{equation}

where 
\begin{equation*}
A\ast _{g}B=\widehat{\alpha }^{-1}\left[ \left( \widehat{\alpha }A\right)
\ast \left( \widehat{\alpha }B\right) \right]
\end{equation*}

\begin{equation}
\widehat{\alpha }:=\alpha \left( -\hbar \frac{\partial ^{2}}{\partial
x\partial p}\right) =f\left( -\hbar \frac{\partial ^{2}}{\partial x\partial p%
}\right) \exp \left\{ \frac{i}{2}\left( -\hbar \frac{\partial ^{2}}{\partial
x\partial p}\right) \right\}  \tag{5}
\end{equation}

and ``$\ast $'' stands for the usual Moyal product

\begin{equation}
A\ast B=A\exp \left\{ \frac{i\hbar }{2}\overleftrightarrow{\mathcal{P}}%
\right\} B  \tag{6}
\end{equation}
\begin{equation*}
A\overleftrightarrow{\mathcal{P}}B:=\left\{ A,B\right\} _{\mathcal{P}}=\frac{%
\partial A}{\partial x}\frac{\partial B}{\partial p}-\frac{\partial A}{%
\partial p}\frac{\partial B}{\partial x}
\end{equation*}
$\blacksquare $
\end{theorem}

It can be also shown that $W_{g}\left( A\right) $ is a symmetric operator
for every real $A\in P[[x,p,\hbar ]]$ if and only if the formal series $%
\alpha =\alpha \left( y\right) =f\left( y\right) \exp \left\{ \frac{i}{2}%
y\right\} $ is real.

In terms of $\alpha $ we have 
\begin{equation}
g\left( m,n,s\right) =\left( \frac{i}{2}\right) ^{s}\frac{m!n!}{\left(
m-s\right) !\left( n-s\right) !}\sum\limits_{k=0}^{s}\frac{\left( 2i\right)
^{k}}{\left( s-k\right) !}\alpha _{k}  \tag{7}
\end{equation}
where $\alpha _{k}$ are defined by 
\begin{equation}
\alpha \left( y\right) =\sum\limits_{k=0}^{\infty }\alpha _{k}y^{k},\quad
\alpha _{0}=1  \tag{8}
\end{equation}
Now we introduce the well known in quantum mechanics operators, $\widehat{a}$
(``the destruction operator'') and its hermitian conjugate $\widehat{a}%
^{\dagger }$ (``the creation operator'') such that 
\begin{equation}
\widehat{x}=\frac{1}{2}\left( \widehat{a}^{\dagger }+\widehat{a}\right)
\qquad \widehat{p}=i\hbar \left( \widehat{a}^{\dagger }-\widehat{a}\right)  
\tag{9}
\end{equation}
\begin{equation*}
\lbrack \widehat{a},\widehat{a}^{\dagger }]=1
\end{equation*}
It is an easy matter to show that 
\begin{equation}
\widehat{x}=\exp \left\{ \frac{1}{2}\left( \widehat{a}^{\dagger }\right)
^{2}\right\} \exp \left\{ \frac{1}{4}\widehat{a}^{2}\right\} \widehat{a}%
^{\dagger }\exp \left\{ -\frac{1}{4}\widehat{a}^{2}\right\} \exp \left\{ -%
\frac{1}{2}\left( \widehat{a}^{\dagger }\right) ^{2}\right\}   \tag{10}
\end{equation}
\begin{equation*}
\widehat{p}=\exp \left\{ \frac{1}{2}\left( \widehat{a}^{\dagger }\right)
^{2}\right\} \exp \left\{ \frac{1}{4}\widehat{a}^{2}\right\} \left( -i\hbar 
\widehat{a}\right) \exp \left\{ -\frac{1}{4}\widehat{a}^{2}\right\} \exp
\left\{ -\frac{1}{2}\left( \widehat{a}^{\dagger }\right) ^{2}\right\} 
\end{equation*}
Therefore one can define an algebra isomorphism 
\begin{equation*}
L:=\widehat{P}[[\widehat{x},\widehat{p},\hbar ]]\longrightarrow \widehat{P}[[%
\widehat{a}^{\dagger },-i\hbar \widehat{a},\hbar ]]
\end{equation*}
by 
\begin{equation}
L\left( \widehat{x}\right) =\widehat{a}^{\dagger }\quad \mathrm{and\quad }%
L\left( \widehat{p}\right) =-i\hbar \widehat{a}.  \tag{11}
\end{equation}
Consequently, by Theorems 1 and 2 we obtain the algebra isomorphism 
\begin{equation*}
L\circ W_{g}:P[[x,p,\hbar ]]\longrightarrow \widehat{P}[[\widehat{a}%
^{\dagger },-i\hbar \widehat{a},\hbar ]]
\end{equation*}
\begin{equation*}
L\circ W_{g}\left( p^{m}x^{n}\right) =\sum\limits_{s=0}^{\min (m,n)}\frac{%
\left( -\hbar \right) ^{s}m!n!}{s!\left( m-s\right) !\left( n-s\right) !}%
\frac{d^{s}f\left( y\right) }{dy^{s}}|_{y=0}\left( -i\hbar \widehat{a}%
\right) ^{m-s}\left( \widehat{a}^{\dagger }\right) ^{n-s}
\end{equation*}
\begin{equation}
\left( L\circ W_{g}\left( A\right) \right) \left( L\circ W_{g}\left(
B\right) \right) =L\circ W_{g}\left( A\ast _{g}B\right) ,\quad A,B\in
P[[x,p,\hbar ]].  \tag{12}
\end{equation}
Now, employing the standard matrix representation of $\widehat{a}$ and $%
\widehat{a}^{\dagger }$ [5,6] 
\begin{equation*}
\widehat{a}\longmapsto a=\left( 
\begin{array}{ccccc}
0 & 1 & 0 & 0 & ... \\ 
0 & 0 & \sqrt{2} & 0 & ... \\ 
0 & 0 & 0 & \sqrt{3} & ... \\ 
\vdots  & \vdots  & \vdots  & \vdots  & 
\end{array}
\right) 
\end{equation*}
\begin{equation}
\widehat{a}^{\dagger }\longmapsto a^{\dagger }=\left( 
\begin{array}{cccc}
0 & 0 & 0 & ... \\ 
1 & 0 & 0 & ... \\ 
0 & \sqrt{2} & 0 & ... \\ 
0 & 0 & \sqrt{3} & 0 \\ 
\vdots  & \vdots  & \vdots  & \vdots 
\end{array}
\right)   \tag{13}
\end{equation}
and substituting the matrices $a$ and $a^{\dagger }$ instead of $\widehat{a}$
and $\widehat{a}^{\dagger },$ respectively, into (12) one finds the algebra
isomorphism $\widetilde{W_{g}}$ between the generalized Moyal algebra $%
\left( P[[x,p,\hbar ]],\ast _{g}\right) $ and the matrix algebra $%
P[[a^{\dagger },-i\hbar a,\hbar ]].$

Denote $F^{\left( m,n\right) }:=\left( -i\hbar a\right) ^{m}\left(
a^{\dagger }\right) ^{n}.$ Simple calculations lead to the following non
vanishing elements of the matrices $F^{\left( m,n\right) }\quad \left(
m+n>0\right) :$%
\begin{equation*}
\left( F^{\left( m,0\right) }\right) _{j,j+m}=\left( -i\hbar \right) ^{m}%
\sqrt{j\left( j+1\right) ...\left( j+m-1\right) },
\end{equation*}
\begin{equation*}
\left( F^{\left( 0,n\right) }\right) _{j+n,j}=\sqrt{j\left( j+1\right)
...\left( j+n-1\right) },
\end{equation*}
\begin{equation*}
\left( F^{\left( m,n\right) }\right) _{j,j+m-n}=\left( -i\hbar \right)
^{m}\left( j+m-n\right) ...\left( j+m-1\right) \sqrt{j\left( j+1\right)
...\left( j+m-n-1\right) },
\end{equation*}
\begin{equation*}
\mathrm{for\quad }m>n>0;
\end{equation*}
\begin{equation*}
\left( F^{\left( m,m\right) }\right) _{j,j}=\left( -i\hbar \right)
^{m}j\left( j+1\right) ...\left( j+m-1\right) ,
\end{equation*}
\begin{equation*}
\left( F^{\left( m,n\right) }\right) _{j+n-m,j}=\left( -i\hbar \right)
^{m}\left( j+n-m\right) ...\left( j+n-1\right) \sqrt{j\left( j+1\right)
...\left( j+n-m-1\right) },
\end{equation*}
\begin{equation}
\mathrm{for\quad }n>m>0.  \tag{14}
\end{equation}
Finally, we have 
\begin{equation}
\widetilde{W_{g}}\left( p^{m}x^{n}\right) =\sum\limits_{s=0}^{\min (m,n)}%
\frac{\left( -\hbar \right) ^{s}m!n!}{s!\left( m-s\right) !\left( n-s\right)
!}\frac{d^{s}f\left( y\right) }{dy^{s}}|_{y=0}F^{\left( m-s,n-s\right) } 
\tag{15}
\end{equation}
This formula corresponds to Merkulov's result but in slightly another
representation and in our case we deal with generalized Moyal products $\ast
_{g}.$

\bigskip

\textbf{Examples}

\bigskip

(1) \textit{The Moyal }$\ast $\textit{-algebra}

It is well known that this algebra is induced by the Weyl ordering of
operators [2,3,4]. In this case the operator $\widehat{\alpha }=1$. Hence by
(5) 
\begin{equation*}
f(y)=\exp \left\{ -\frac{i}{2}y\right\} \Longrightarrow \frac{d^{s}f(y)}{%
dy^{s}}|_{y=0}=\left( -\frac{i}{2}\right) ^{s}
\end{equation*}
and we get now (the index ``g'' is omitted) 
\begin{equation}
\widetilde{W}\left( p^{m}x^{n}\right) =\sum\limits_{s=0}^{\min (m,n)}\frac{%
\left( i\hbar \right) ^{s}m!n!}{2^{s}s!\left( m-s\right) !\left( n-s\right) !%
}F^{\left( m-s,n-s\right) }  \tag{16}
\end{equation}
(compare with Merkulov's result)

\bigskip

(2) \textit{The }$\ast _{(st)}$\textit{-algebra}

This algebra follows from the standard ordering 
\begin{equation*}
p^{m}x^{n}\longmapsto \widehat{x}^{n}\widehat{p}^{m}
\end{equation*}
Here $\alpha (y)=\exp \left\{ -\frac{i}{2}y\right\} .$ Hence 
\begin{equation*}
f(y)=\exp \left\{ -iy\right\} \Longrightarrow \frac{d^{s}f(y)}{dy^{s}}%
|_{y=0}=\left( -i\right) ^{s}
\end{equation*}
Consequently 
\begin{equation}
\widetilde{W}_{st}\left( p^{m}x^{n}\right) =\sum\limits_{s=0}^{\min (m,n)}%
\frac{\left( i\hbar \right) ^{s}m!n!}{s!\left( m-s\right) !\left( n-s\right)
!}F^{\left( m-s,n-s\right) }  \tag{17}
\end{equation}

\bigskip

(3) \textit{The }$\ast _{(ast)}$\textit{-algebra}

This is the algebra which follows from the anti-standard ordering 
\begin{equation*}
p^{m}x^{n}\longmapsto \widehat{p}^{m}\widehat{x}^{n}
\end{equation*}
Now $\alpha (y)=\exp \left\{ \frac{i}{2}y\right\} .$ Hence $f(y)=1$ and it
remains only one term with $s=0$ in (15).

Hence 
\begin{equation}
\widetilde{W}_{ast}\left( p^{m}x^{n}\right) =F^{\left( m,n\right) }  \tag{18}
\end{equation}
(Compare with Merkulov's paper [1]).

\bigskip

(4) \textit{The }$\ast _{(sym)}$\textit{-algebra}

Here we deal with the algebra generated by the symmetric ordering. So one
has $\alpha (y)=\cos \left( \frac{y}{2}\right) .$ Therefore, 
\begin{equation*}
f(y)=\frac{1}{2}\left( 1+\exp \left\{ -iy\right\} \right) \Longrightarrow 
\frac{d^{s}f(y)}{dy^{s}}|_{y=0}=\frac{1}{2}\left( \delta _{s,0}+\left(
-i\right) ^{s}\right) .
\end{equation*}
Consequently: 
\begin{equation}
\widetilde{W}_{sym}\left( p^{m}x^{n}\right) =F^{\left( m,n\right)
}+\sum\limits_{s=1}^{\min (m,n)}\frac{\left( i\hbar \right) ^{s}m!n!}{%
2(s!)\left( m-s\right) !\left( n-s\right) !}F^{\left( m-s,n-s\right) } 
\tag{19}
\end{equation}
\bigskip Finally we consider

\bigskip

(5) \textit{The }$\ast _{BJ}$\textit{-algebra}

This algebra follows from the Born-Jordan ordering.

Now $\alpha (y)=\frac{\sin \left( \frac{y}{2}\right) }{\left( \frac{y}{2}%
\right) }.$ Therefore 
\begin{equation*}
f(y)=\frac{1}{iy}\left( 1-\exp \left\{ -iy\right\} \right) \Longrightarrow 
\frac{d^{s}f(y)}{dy^{s}}|_{y=0}=\frac{\left( -i\right) ^{s}}{s+1}
\end{equation*}
Hence 
\begin{equation*}
\widetilde{W}_{BJ}\left( p^{m}x^{n}\right) =\sum\limits_{s=0}^{\min (m,n)}%
\frac{\left( i\hbar \right) ^{s}m!n!}{(s+1)!\left( m-s\right) !\left(
n-s\right) !}F^{\left( m-s,n-s\right) }.
\end{equation*}

\bigskip

\bigskip

\textbf{Acknowledgments}

We are indebted to Hugo Garc\'{i}a-Compe\'{a}n for pointing out Merkulov's
paper. This paper is partially supported by CONACYT and CINVESTAV
(M\'{e}xico) and by KBN (Poland). M. Przanowski thanks the staff of
Departamento de F\'{i}sica at CINVESTAV, (M\'{e}xico, D.F.) for warm
hospitality.


\newpage

\begin{thebibliography}{9}
\bibitem{Merkulov}  S.A. Merkulov, ''The Moyal product is the matrix
product'' arXiv:math-ph/0001039 V2 31 Jan 2000.

\bibitem{Wolf}  K.B. Wolf, \textit{The Heisenberg-Weyl Ring in Quantum
Mechanics}, in \textit{Group Theory and Its Application}, ed. E. Loebl
(Academic Press, New York 1975) Vol. III, p.p. 189-247.

\bibitem{Tosiek}  J. Tosiek and M. Przanowski, Acta Phys. Pol. \textbf{B 26}%
, 1703 (1995).

\bibitem{Pleb}  J.F. Pleba\'{n}ski, M. Przanowski and J. Tosiek, Acta Phys.
Pol. \textbf{B 27}, 1961 (1996).

\bibitem{Schiff}  L.I. Schiff, \textit{Quantum Mechanics} (McGraw-Hill,
Inc., 1968).

\bibitem{Messiah}  A. Messiah, \textit{Quantum Mechanics}, Vol. 1
(North-Holland, Amsterdam, 1961).
\end{thebibliography}
\end{document}